\documentclass[twocolumn, pra,showpacs,superscriptaddress]{revtex4}
\usepackage{amssymb}
\usepackage{mathrsfs}
\usepackage{graphicx}
\usepackage{subfigure}
\usepackage{float}
\usepackage[english]{babel}
\usepackage[utf8x]{inputenc}
\usepackage[T1]{fontenc}
\usepackage{color}
\usepackage{CJK}
\usepackage{natbib}
\usepackage{dcolumn}
\usepackage{bm}
\usepackage{amsmath}
\newlength{\figurewidth}
\begin{document}

\preprint{APS/123-QED}

\title{Correlation in Momentum Space of Tonks-Girardeau Gas}
\author{Yajiang Hao}
\email{haoyj@ustb.edu.cn}
\author{Yiwang Liu}
\affiliation{Institute of Theoretical Physics and Department of Physics, University of Science and Technology Beijing, Beijing 100083, China}

\author{Xiangguo Yin}
\affiliation{Institute of Theoretical Physics and State Key Laboratory of Quantum Optics and Quantum Optics Devices, Shanxi University, Taiyuan 030006, China}
\date{\today}

\begin{abstract}
We investigate the correlation properties of the ground state of Tonks-Gigrardeal gases in the momentum space. With Bose-Fermi mapping method the exact ground state wavefunction in coordinate space can be obtained basing on the wavefunction of spin-polarized Fermions. By Fourier transformation we obtain the ground state wavefunction in momentum space, and therefore the momentum distribution, pair correlation and the reduced one-body density matrix (ROBDM) in momentum space. The ROBDM in momentum space is the Fourier transformation of the ROBDM in coordinate space and the pair correlation in momentun space is the Fourier transformation of the reduced two-body density matrix in coordinate space. The correlations in momentum space display larger values only in small momentum region and vanish in most other regions. The lowest natural orbital and occupation distribution are also obtained.
\end{abstract}
\maketitle

\section{Introduction}
Quantum correlation is important in many-body quantum system for its close relations with physical and chemical properties of system. Particularly, quantum correlation is usually related with quantum many-particle interference and coherence, and even the specific quantum phase. The dependence of correlation effects on the internal parameters and external parameters has always been the research focus. Besides the traditional condensed matter system, the neutral ultracold atomic system has become a crucial platform to investigate quantum many body correlation for its high controllability of the atomic interaction and dimension. With the Feschbach resonance and confined induced resonance technique \cite{FR,CIR} the atomic interaction can be tuned such that the system might in the weak interaction regime or in the strong interaction regime. With optical lattices or the highly anisotropic trap the three-dimensional ultracold atom system can become an effective one-dimensional (1D) or two-dimensional quantum system \cite{Paredes,Toshiya,Ketterle,Single1D}. Both the interaction and dimensionality play critical roles in the correlation of quantum many-body system \cite{RMP2011,RMP2012,RMP2013}. The $M$-order correlation of cold atom system, including low-dimensional quantum gas, have attract many interests of theorists \cite{Sykes,Tolra,Kinoshita,Brandt,Sekino} and experimentalists \cite{nbodycorr2013,corrlongrange}. The dependence on temperature and time is also the important theme \cite{Kheruntsyan,Kozlowski,Patu,Gamayun}.

 Besides the interaction, the reduction of dimensions can also affect the quantum correlation. Tonks-Gigrardeau (TG) gas \cite{TG1,TG2} is a strong-correlation system, which is the 1D ultracold bosonic atom gases with infinite repulsive interaction. The one-body, two-body and three-body correlation functions \cite{Devillard,Olshanii2017}, static and dynamical structure factor\cite{SSF,DSF}, and universal contact of the system has been studied\cite{WXu}. So far, most related study are done in coordinate space. But with the development of measurement technique such as momentum microscope \cite{Hodgman} and single-atom-resolved detection methods \cite{Cayla,HOtt} quantum correlation in momentum space is becoming a popular theme \cite{Bergschneider,Brandt2018,Butera,Preiss,Becher,Carcy}. The investigation of quantum correlation, particularly higher-order correlation, in momentum space is an important issue and greatly desirable for characterization of quantum phase, quantum-information processes and quantum simulation in 1D system \cite{BFang}.

In the present paper we will investigate the correlation properties of ground state of TG gas. Theoretically the exact many-body wavefunction in coordinate space of TG gases can be obtained by Bose-Fermi mapping method basing on the Slater-type ground-state wavefunction of the 1D polarized Fermi gas. While the wavefunction in momentum space is the Fourier transformation of the wavefunction in coordinate space. Using the wavefunction we get the quantum correlation in momentum space by integrating on partial particle freedom. It turns out that the correlation function in momentum space are Fourier transformation of correlation function in coordinate space. The reduced one-body density matrix and thus the density distribution in momentum space are the Fourier transformation of the reduced one-body density matrix in coordinate space. The pair correlation in momentum space is the Fourier transformation of reduced two-body density matrix. The natural orbital and the corresponding occupation number in momentum space can also be obtained by diagonalizing the reduced one-body density matrix in momentum space \cite{Yannouleas,BFang}.

The paper is organized as follows. In Sec. II, we obtain the exact ground state wavefunction in coordinate space of TG gas with Bose-Fermi mapping method. In Sec. III, we present the density distribution, pair correlation, the reduced one-body density matrix, occupation distribution and the lowest natural orbitals in momentum space. A brief summary is given in Sec. IV.

\section{Ground state wavefunction}

The 1D Bose gas composed of $N$ atoms with mass $m$ confined in an 1D trap can be described by the Hamiltonian
\begin{equation}
H=\sum_{j=1}^N-\frac{\hbar ^2}{2m}\frac{\partial ^2}{\partial x_i^2}+g_{1D}\sum_{j<l}\delta(x_j-x_l),
\end{equation}
with $g_{1D}$ being the effective 1D interaction strength. The many-body wavefunction $\Psi(x_{1},\cdots ,x_{N})$ satisfy the Schr\"{o}dinger equation $H\Psi(x_{1},\cdots ,x_{N})=E\Psi(x_{1},\cdots ,x_{N})$. For the strongly interacting TG gas $g_{1D}\rightarrow +\infty $ such that the wavefunction is constrained by the boundary condition
\begin{equation*}
\Psi(x_{1},\cdots ,x_{N})=0 \text{ for } x_j=x_l,
\end{equation*}
which is equivalent to the constraint on wavefunction of the polarized fermions. Therefore the exact wavefunction of TG gas confined in a box of length $L$ can be obtained by Bose-Fermi mapping method based on the wavefunction of polarized fermions
\begin{align}
\Psi_F (x_{1},\cdots ,x_{N}) =(L^{N}N!)^{-1/2}\det
[e^{ik_{l}x_{j}}]_{l=1,...,N}^{j=1,...,N}
\end{align}
with $k_{l} =2\pi /L(l-(N+1)/2)$. Here we assume that the system is confined on a circle of circumference length $L$ and the wavefunction satisfy periodical boundary condition. For simplify the atom number is assumed to be odd. Using the Vandermonde determinant formula
\begin{equation}
\det [p_{j-1}(x_{k})]_{j,k=1,\cdots ,N}=\prod_{1\leq j<k\leq N}(x_{j}-x_{k})
\end{equation}
for $\{p_{j}(x)=x^{j}\}$, we reformulate the Fermi wavefunction of $N$ particles as
\begin{align}
\Psi_F (x_{1},\cdots ,x_{N})=(L^{N}N!)^{-1/2}\prod_{j=1,...,N}e^{-i(N-1)\pi
x_{j}/L}  \nonumber \\
\times \prod_{1\leq j<l\leq N}(e^{i2\pi x_{j}/L}-e^{i2\pi x_{l}/L}).
\end{align}
With the Bose-Fermi mapping method we can obtain the exact ground state wave function of TG, which is symmetric under exchange of two atoms
\begin{align}
\Psi\left( x_{1},x_{2},\cdots ,x_{N}\right)
=(L^{N}N!)^{-1/2}\prod_{j=1,...,N}e^{-i(N-1)\pi x_{j}/L}  \nonumber  \\
\times \prod_{1\leq j<l\leq N}\epsilon (x_j-x_l)\left( e^{i2\pi x_{j}/L}-e^{i2\pi x_{l}/L}\right),     \label{WFx}
\end{align}
where the sign function $\epsilon(x_j-x_l)$ are 0, +1, and -1 for $x_j-x_l$=0, $>0$ and $<0$, respectively. Here $\epsilon(x_j-x_l)$ are introduced to ensure that $\Psi\left( x_{1},x_{2},\cdots ,x_{N}\right)$ satisfy the exchange symmetry of identical bosons.

\section{Momentum distribution and correlation function in momentum space}
The wavefunction in momentum space $\Psi (k_{1},\cdots,k_N)$ is related to the wavefunction in coordinate space $\Psi (x_{1},\cdots ,x_{N})$ by the following Fourier transformation
\begin{align}
\Psi (k_{1},\cdots ,k_{N-1},k_N)=\frac{1}{\left( 2\pi \right) ^{N/2}}
\int_{0}^{L}dx_{1}\cdots dx_{N} \nonumber \\
\times \exp \left[-i\sum_{j=1}^{N}k_{j}x_{j}\right] \Psi (x_{1},\cdots ,x_{N}).  \label{WFk}
\end{align}

For simplicity we take the length unit as $\frac{L}{2\pi}$ and momentum unit as $\frac{2\pi}{L}$. In the following section all quantities are dimensionless and we preserve the original notation.
\begin{figure}
    \centering
    \includegraphics[width=3.4in]{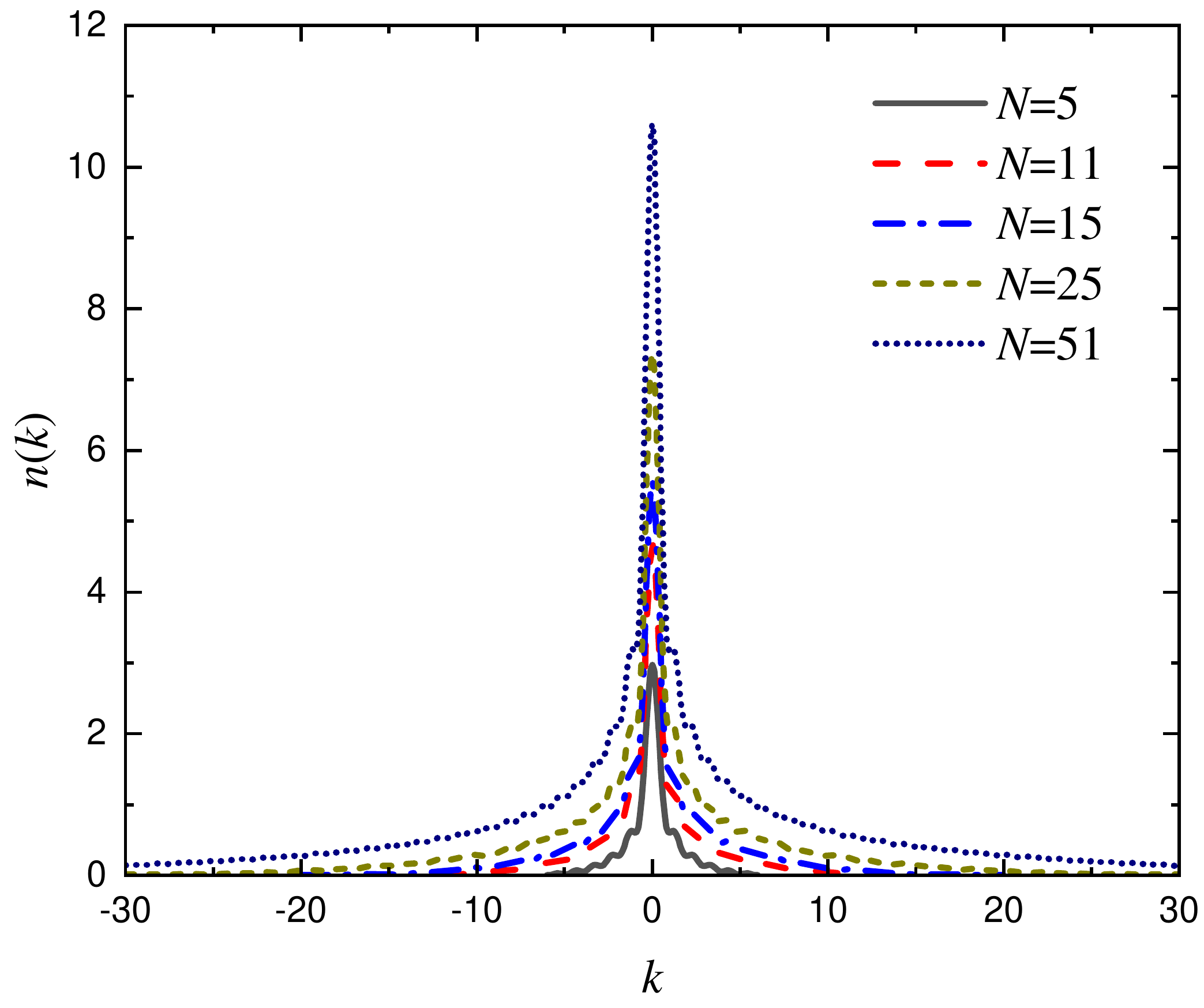}
    \caption{Momentum distribution of TG gas for different atom number. $k$ is in unit of $2\pi/L$ and $n(k)$ is in unit of $L/2\pi$.}
    \label{fig:MomDis}
\end{figure}

The momentum distribution can be obtained by integrating on $N-1$ particle coordinates
\begin{align}
n(k)& =N\int_{-\infty}^{\infty}dk_{1}\cdots \int_{-\infty}^{\infty}dk_{N-1}   \\
& \times \Psi _{A}^{\ast}(k_{1},\cdots ,k_{N-1},k_{N})\Psi _{A}(k_{1},\cdots ,k_{N-1},k_{N}). \nonumber
\end{align}
Introducing Eq. (\ref{WFx}) and Eq. (\ref{WFk}) into Eq. (7) it turns out that the momentum distribution is the Fourier transformation of reduced one-body density matrix in coordinate space (ROBDM-CS) $\rho(x,y)$
\begin{align*}
n(k)=(2\pi)^{-1}\int dxdy\rho(x,y) \exp \left[ ik\left(x-y\right) \right]
\end{align*}
with the ROBDM-CS (Appendix A) being (in unit of $2\pi/L$)
\begin{equation*}
\rho(x,y)=\frac{1}{(2\pi )^{N}}e^{i(N-1)(x-y)/2}\det [b_{j,k}\left(
x,y\right) ]_{j,k=1,...,N-1},
\end{equation*}
which is the probability to find the particles at two positions $x$ and $y$ in two successive measurements, respectively. The diagonal part of $\rho(x,y)$ is density distribution in coordinate space $\rho (x)=\rho(x,y)=\frac{N}{L}$, which shows that the strongly interacting Bose atoms distribute homogeneously in the whole coordinate space. It is greatly different from the distribution in the momentum space, which is a single-peak structure. We plot the momentum distribution of TG gases for different atoms number $N$ in Fig. 1. It is shown that the strongly interacting Bose atoms mainly distribute in the region of small momentum as if atoms condensate in the momentum space although it is not a true Bose-Einstein condensate.

\begin{figure}
    \centering
    \includegraphics[width=3.4in]{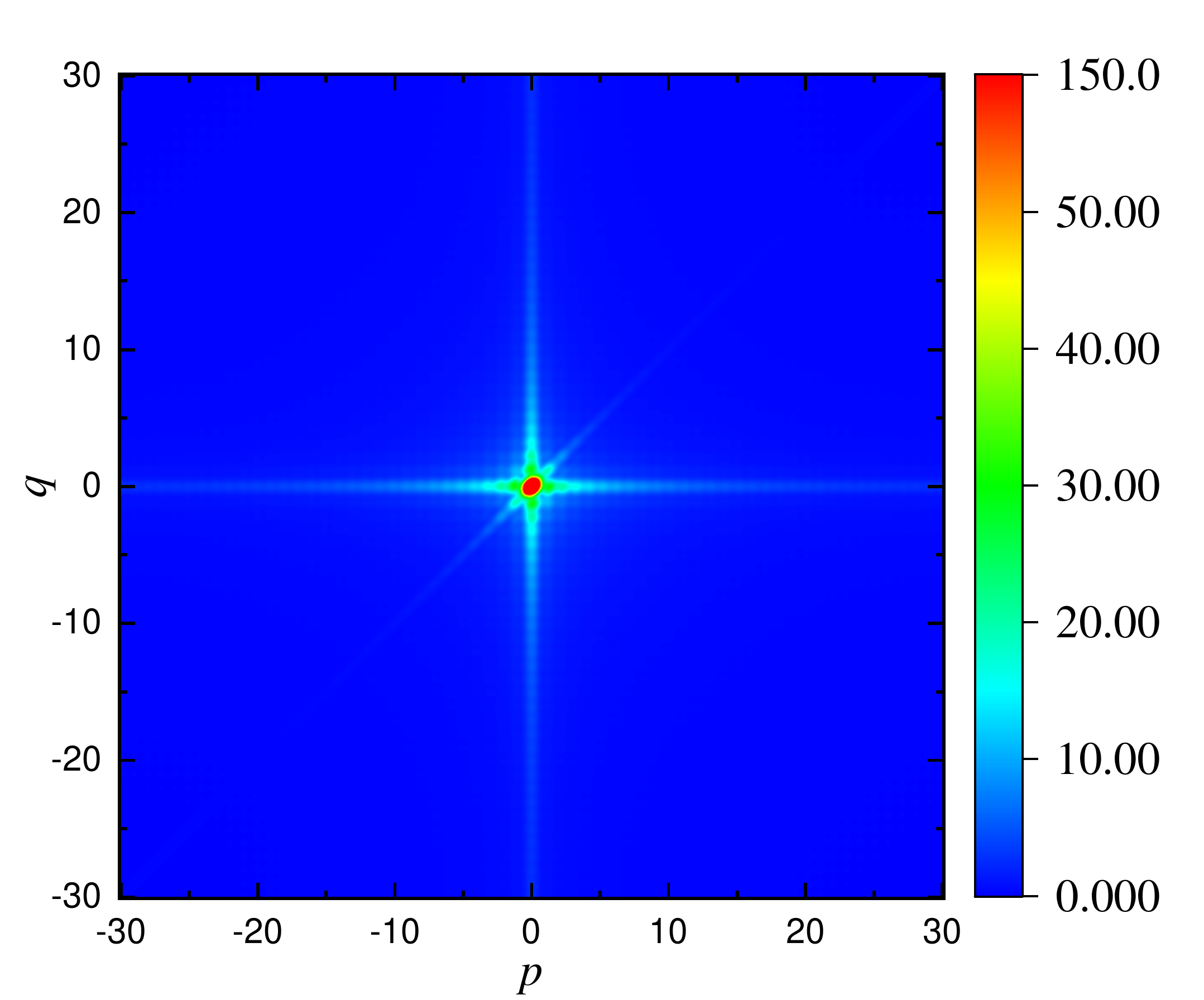}
    \caption{Pair correlation in momentum representation for TG gas with $N=51$. $p$ and $q$ are in unit of $2\pi/L$ and $\rho(p,q)$ is in unit of $(L/2\pi)^2$.}
    \label{fig:pc}
\end{figure}

The pair correlation denotes the joint probability in two successive measurement where the first measurement find one atom with momentum $p$ and the immediate second measurement find one atom with momentum $q$. In momentum space it is defined as
\begin{align}
\rho (p,q)& =N(N-1)\int_{-\infty }^{\infty }dk_{1}\cdots \int_{-\infty
}^{\infty }dk_{N-2}  \nonumber      \\
\times &  \left |\Psi(k_{1},\cdots,k_{N-2},p,q)\right |^2
\end{align}%
Using Eq. (\ref{WFk}) we get its dependence on the reduced two-body density matrix $\rho_2(x,y;x',y')$ in coordinate space (in unit of $(L/2\pi)^2$)
\begin{align}
\rho (p,q)&=\frac{1}{\left( 2\pi \right) ^{2}}\int_{0}^{2\pi}dxdy
\int_{0}^{2\pi}dx^{\prime }dy^{\prime }        \\
\times &\exp \left( ip\left( x-x^{\prime
}\right) \right) \exp \left( iq\left( y-y^{\prime }\right) \right) \rho_2(x,y,x^{\prime },y^{\prime }).       \nonumber
\end{align}

With the obvious expression of wavefunction in coordinate space Eq. (\ref{WFx}), the two-body density matrix (in unit of $(2\pi/L)^2$) is formulated as (Appendix B)
\begin{align}
\rho_2(x,y;x^{\prime },y^{\prime }) &=\frac{1}{(2\pi )^{N}}%
e^{i(N-1)(x+y-x^{\prime }-y^{\prime })/2}\epsilon (x-y) \nonumber \\
&\times \epsilon (x^{\prime
}-y^{\prime })(e^{-ix}-e^{-iy})(e^{ix^{\prime }}-e^{iy^{\prime }}) \nonumber  \\
&\times \det [b_{j-k}(x,y;x^{\prime },y^{\prime })]_{j,k=1,...,N-2}
\end{align}

The pair correlation in momentum space of TG gas for $N=51$ is plotted in Fig. 2. It is shown that in most regions the pair correlation vanishes except the zero momentum region, which shows that two atoms with strong interaction behave at the least momentum simultaneously with the extremely largest probability. This is consistent with the statistical properties of bosons that bosons prefer to condensates in the same quantum state even though the Bose atoms repel each other with the infinite interaction in coordinate space. Furthermore, there are finite probabilities with which two atoms behave at different momentum but the probability is extremely small and at least one atom is at the small momentum region (two momentum axes). It is interesting to notice that there are finite probability that two atoms behave at same momentum (the diagonal part) but the probability that two atoms behave at opposite momentum is zero (the anti-diagonal part) except that the much small momentum region.

\begin{figure}
    \centering
    \includegraphics[width=3.4in]{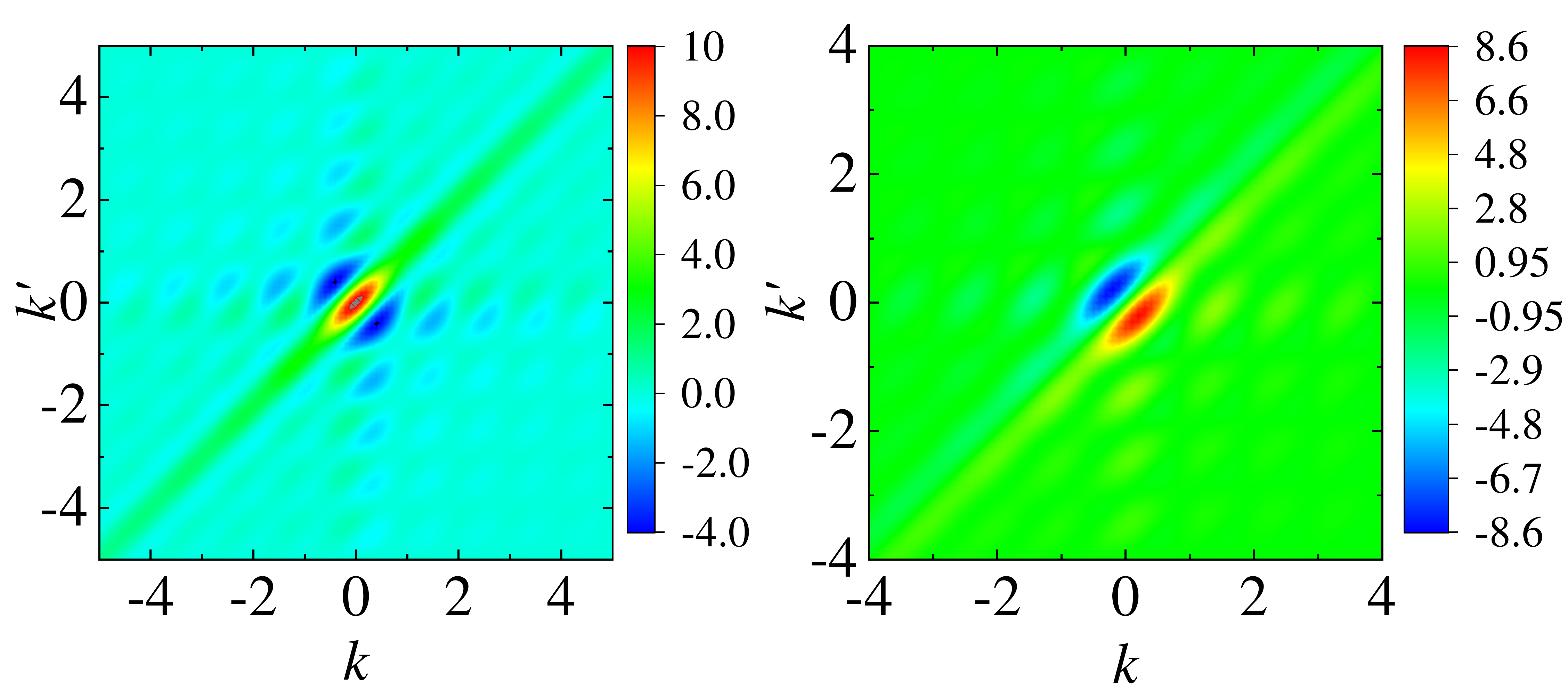}
    \caption{ROBDM in momentum space for TG gas with $N=51$. Left: The real part of ROBDM-MS; Right: The imaginary part of ROBDM-MS.}
    \label{fig:ROBDM-MS}
\end{figure}

The reduced one-body density matrix in momentum space (ROBDM-MS) denote the probability to find the particles behaving with momentum $k$ and $k^{\prime}$ in two successive measurements, respectively. It is defined as
\begin{align*}
\rho(k,k')& =N\int dk_{1}\cdots \int dk_{N-1} \\
\times & \Psi ^{\ast }(k_{1},\cdots ,k_{N-1},k)\Psi (k_{1},\cdots ,k_{N-1},k').
\end{align*}
With the same procedure as the above we have
\begin{equation}
\rho(k,k^{\prime})=\frac{1}{2\pi }\int_{0}^{2\pi}dxdy\exp i\left(kx-k' y\right) \rho(x,y),
\end{equation}
which is the Fourier transformation of ROBDM-CS $\rho (x,y)$.
Its diagonal part is momentum distribution, i.e., $\rho(k,k)=n(k)$. We display the ROBDM-MS in Fig. 3 for TG gas with $N=51$. It is shown that in most regions of $(k,k')$-plane $\rho(k,k^{\prime})$ is zero except that the small momentum region and the ROBDM-MS is a diagonal dominant complex conjugate matrix. Therefore in two successive measurements the probability to find two atoms with almost same momentum is the largest. Along the region nearby the momentum axes, the ROBDM-MS decrease osciallately. This is extremely different from the properties of ROBDM-CS, which depends only $|x-y|$ and $\rho (x,y)=\rho(|x-y|)$. It is a diagonal dominant real symmetry matrix and decreases monotonously to the least value at $\pi/2$, and then increase  with the increase of $|x-y|$ (Fig. 5 in Appendix A).

\begin{figure}
    \centering
    \includegraphics[width=3.4in]{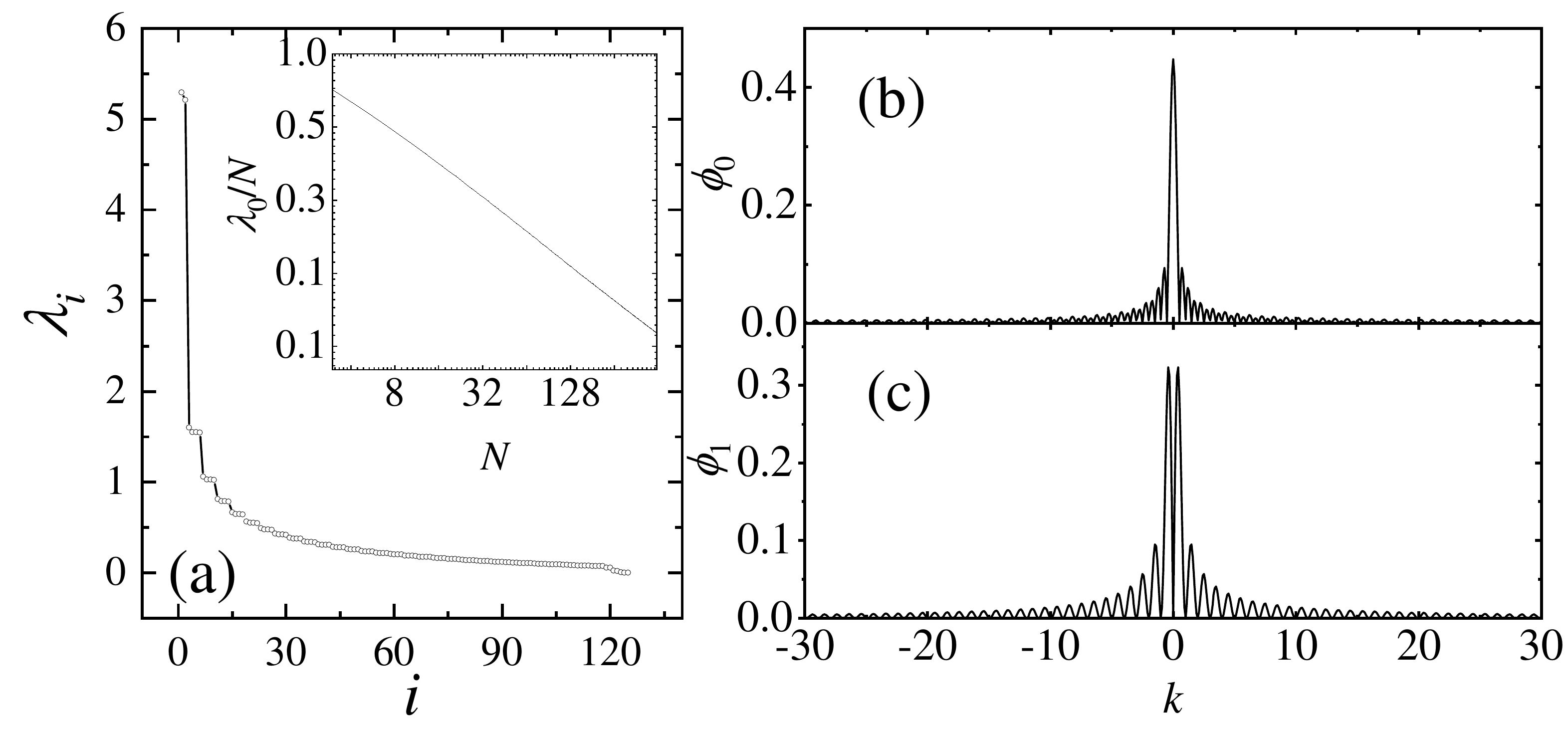}
    \caption{Natural orbital and occupation number in momentum space for TG gas. (a) Occupation number for $N=51$; (b) The lowest natural orbital for $N=51$. (c) The second lowest natural orbital for $N=51$. Inset: The fractional occupation of the lowest natural orbital $\lambda_0/N$ versus the total atom atom number $N$.}
    \label{fig:NatOrb}
\end{figure}
By solving the eigen equation we have the natural orbital $\phi_j(k)$ and the correspondent occupation $\lambda_j$
\begin{align*}
\int \rho(k,k^{\prime})\phi_j(k^{\prime})dk^{\prime} =\lambda_j\phi_j(k),
\end{align*}
where $\lambda_j$ is the occupation number of the natural orbital $\phi_j(k)$ and $\sum_j \lambda_j=N$. In Fig. 4(a) we show the occupation distribution for different natural orbital of TG gas with $N=51$. The natural orbitals are ordered according to the magnitude of eigenvalues. In the inset of Fig. 4 we plot the log-log plot of fractional occupation of the lowest orbital $\lambda_0/N$ versus the atom number $N$. It is shown that the fractional occupation decrease with the increase of $N$. The lowest and second-lowest natural orbitals are also displayed in Fig. 4 (b) and in Fig. 4 (c). It is shown that they are local and both distribute in the small momentum region, which is extremely different from the natural orbitals in the coordinate space that are nonlocal and distribute in the whole coordinate space \cite{Girardeau2001}.

\section{Conclusion}
In conclusion, we investigated the quantum correlation of the ground state of the TG gas in the momentum space. With Bose-Fermi mapping method we obtain the exact ground state wavefunction in coordinate space. While the many body wavefunction in momentum space is its Fourier transformation. It is shown that the ROBDM-MS and momentum distribution are the Fourier transformation of ROBDM-CS, pair correlation is Fourier transformation of the reduced two-body density matrix in coordinate space.

It turns out that the ground state of TG gas exhibit extremely different properties in momentum space from those in coordinate space. For the TG gas confined in a homogeneous potential, the Bose atom distribute in the whole coordinate space and $\rho(x)=\frac{N}{L}$ while in momentum space atoms occupy only in a small momentum region nearby the zero momentum. The pair correlation in momentum space also exhibit that two atoms prefer to occupy the zero momentum region simultaneously with the extremely large probability. In most momentum region the pair correlation vanish except that with the finite probability two atoms have same momentum or at least one of them has very small momentum. The ROBDM-MS is a diagonal dominant complex conjugate matrix while the ROBDM-CS is a real symmetry matrix dependent only on $|x-y|$. The natural orbitals in momentum space also display different behaviour from those in coordinate space. The former distribute in the small momentum region locally and the later always distribute in the whole coordinate space.

\begin{acknowledgments}
This work was supported by NSF of China under Grants No. 11774026.
\end{acknowledgments}

\appendix
\section{ROBDM in coordinate space}
\begin{figure}
    \centering
    \includegraphics[width=3in]{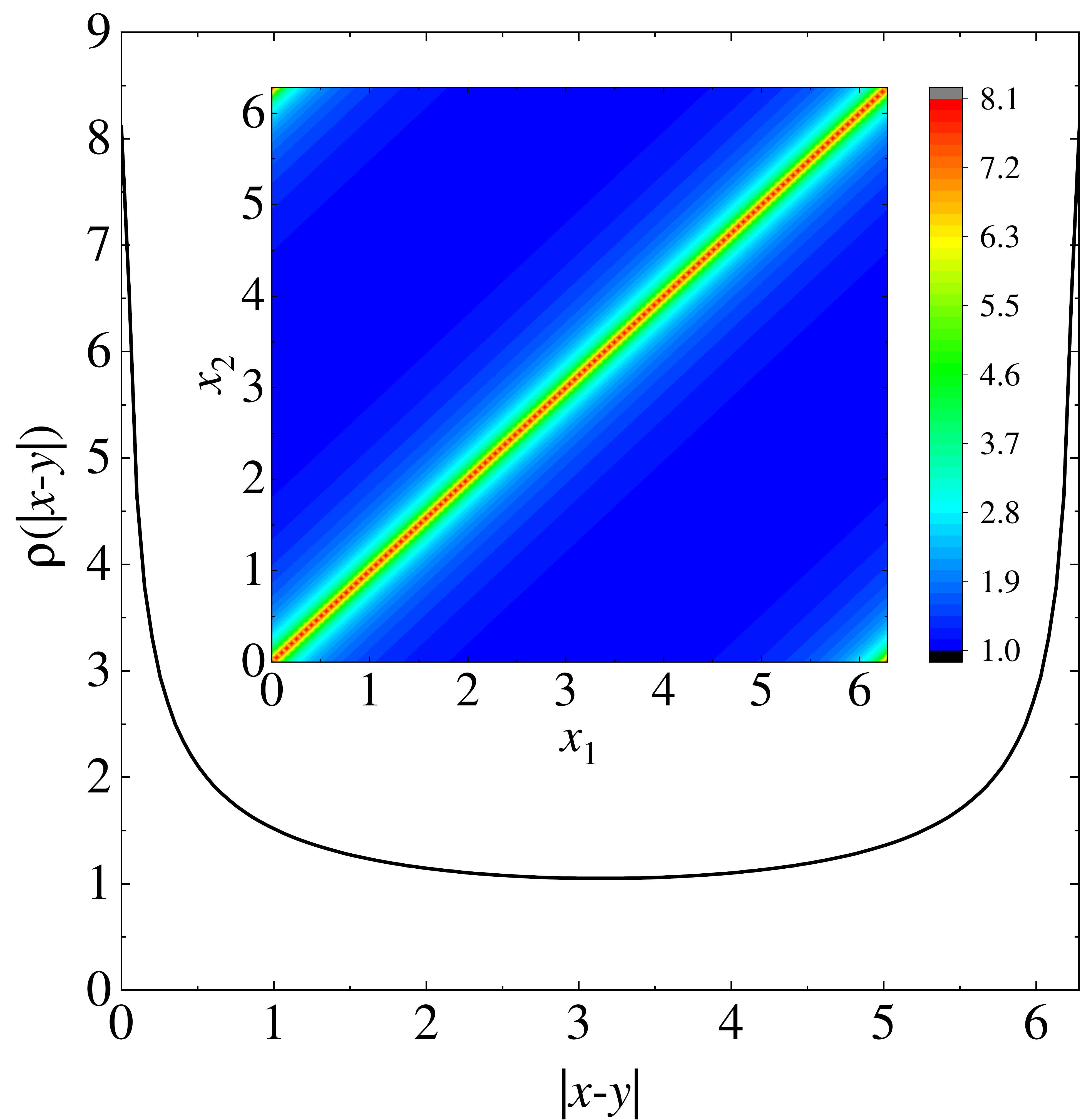}
    \caption{The reduced one-body density matrix $\rho(|x-y|)$ in coordinate space for TG gas with $N=51$. Inset: The ROBDM-CS $\rho(x,y)$.}
    \label{fig:ROBDM-CS}
\end{figure}
The ROBDM in coordinate space is
\begin{align}
\rho(x,y)& =N\int_{0}^{L}dx_{1}\cdots \int_{0}^{L}dx_{N-1} \nonumber \\
\times & \Psi^{\ast }(x_{1},\cdots ,x_{N-1},x)\Psi (x_{1},\cdots,x_{N-1},y).
\end{align}
With Eq. (\ref{WFx}) we have (in unit of $2\pi /L$)
\begin{align*}
\rho(x,y)& =\frac{1}{(N)!(2\pi )^{N-1}}e^{i(N-1)(x-y)/2}%
\prod_{l=1}^{N-1}\int_{0}^{2\pi }dx_{l}  \\
& \times \prod_{1\leq l\leq N-1}\epsilon
(x_{l}-x)\epsilon (x_{l}-y)\left( e^{-ix_{l}}-e^{-ix}\right) \left(
e^{ix_{l}}-e^{iy}\right)  \\
& \times \prod_{1\leq j<k\leq N-1}|e^{ix_{j}}-e^{ix_{l}}|^{2}.
\end{align*}
Utilizing the following result of Toeplitz matrix \cite{Lenard}
\begin{align}
& \frac{1}{N!}\prod_{l=1}^{N}\int_{0}^{2\pi }dx_{l}g(x_{l})\prod_{1\leq
j<k\leq N}|e^{ix_{j}}-e^{ix_{k}}|^{2}   \nonumber \\
& =\det [\int_{0}^{2\pi }dtg(t)\exp (it(j-k))]_{j,k=1,...,N},
\end{align}
we have the ROBDM-CS
\begin{equation}
\rho(x,y)=\frac{1}{(2\pi )^{N}}e^{i(N-1)(x-y)/2}\det [b_{j,k}\left(x,y\right) ]_{j,k=1,...,N-1}
\end{equation}
with the matrix element being
\begin{align*}
b_{j,k}\left( x,y\right)& =\int_{0}^{2\pi }dt\epsilon (t-x)\epsilon(t-y)  \nonumber \\
&\times \left( e^{-it}-e^{-ix}\right) \left( e^{it}-e^{iy}\right) e^{it(j-k)}.
\end{align*}

To simplify the calculation, we reformulate $b_{j,k}(x,y)$ as
\begin{equation*}
b_{j,k}(x,y)=f_{j,k}(x,y)-2\epsilon (y-x)g_{j,k}(x,y),
\end{equation*}
in which $f_{j,k}(x,y)$ and $g_{j,k}(x,y)$ are introduced as following
\begin{align*}
f_{j,k}(x,y) &= \left( 1+e^{i\left( y-x\right) }\right) 2\pi \delta_{j,k}  \\
&-e^{iy}2\pi \delta _{j,k+1}-e^{-ix}2\pi \delta _{j,k-1},
\end{align*}
and
\begin{align*}
g_{j,k}(x,y)&=(1+e^{-i(x-y)})\mu _{j-k}(x,y)\\
& -e^{iy}\mu_{j-k-1}(x,y)  -e^{-ix}\mu _{j-k+1}(x,y)
\end{align*}%
with $\mu _{m}(x,y)$ being $y-x$ for $m=0$ and $\mu _{m}(x,y)$ being $-\frac{i}{m}(e^{imy}-e^{imx})$ for $m\neq 0$.

The diagonal part $\rho(x,x)$ is the density distribution and we reformulate it as $\rho(x)$. With Eq. (A3) we have
\begin{equation*}
\rho(x)=\frac{1}{(2\pi )^{N}}\det [b_{j,k}\left( x,y\right)
]_{j,k=1,...,N-1}
\end{equation*}
The simple evaluation shows that $b_{j,k}=4\pi$, $-2\pi e^{\pm ix}$ for $j-k=0$ and $\pm 1$, respectively, and $b_{j,k}=0$ for others. Therefore we have the homogeneous density distribution  $\rho(x)=\frac{N}{2\pi} $. As the unit is included we have $\rho(x)=\frac{N}{L} $. The ROBDM-CS of TG gas with $N=51$ is displayed in Fig. 5. It shows that ROBDM-CS is diagonal dominant real symmetry, and $\rho(x,y)=\rho(|x-y|)$.

\section{Two-body density matrix in coordinate space}

The two-body density matrix in coordinate space is
\begin{align*}
\rho_2(x,y;x^{\prime },y^{\prime })& =N(N-1)\int_{-\infty }^{\infty
}dx_{1}\cdots \int_{-\infty }^{\infty }dx_{N-2} \\
\times & \Psi^{\ast }(x_{1},\cdots ,x_{N-2},x,y)\Psi(x_{1},\cdots
,x_{N-2},x^{\prime },y^{\prime }).   \nonumber
\end{align*}
Introducing the wavefunction Eq. (\ref{WFx}) and with the length unit given in the main text we have $\rho_2(x,y;x^{\prime },y^{\prime }) $ (in unit of $(2\pi/L)^2$)
\begin{align}
\rho_2(x,y;x^{\prime },y^{\prime }) &=\frac{1}{(2\pi )^{N}}%
e^{i(N-1)(x+y-x^{\prime }-y^{\prime })/2}\epsilon (x-y) \nonumber \\
&\times \epsilon (x^{\prime
}-y^{\prime })(e^{-ix}-e^{-iy})(e^{ix^{\prime }}-e^{iy^{\prime }})  \nonumber \nonumber\\
&\times \det [b_{j-k}(x,y;x^{\prime },y^{\prime })]_{j,k=1,...,N-2}
\end{align}
with the matrix element
\begin{align*}
b_{j-k}(x,y;x^{\prime },y^{\prime })& =\int_{0}^{2\pi }dt\epsilon(t-x)\epsilon (t-y)   \\
&\times \epsilon (t-x^{\prime })\epsilon (t-y^{\prime
})f(x,y;x^{\prime },y^{\prime },t),
\end{align*}
in which $f(x,y;x^{\prime },y^{\prime })$ is introduced
\begin{align*}
f(x,y;x^{\prime },y^{\prime })&
=(e^{-it}-e^{-ix})(e^{-it}-e^{-iy}) \\
& \times (e^{it}-e^{ix^{\prime
}})(e^{it}-e^{iy^{\prime }})\exp (it(j-k)).
\end{align*}

\end{document}